# Two-Dimensional Gallium Sulfide Nanoflakes for UV-Selective Photoelectrochemical-type Photodetectors

Marilena I. Zappia, Gabriele Bianca, Sebastiano Bellani,* Nicola Curreli, Zdeněk Sofer, Michele Serri, Leyla Najafi, Marco Piccinni, Reinier Oropesa-Nuñez, Petr Marvan, Vittorio Pellegrini, Ilka Kriegel, Mirko Prato, Anna Cupolillo, and Francesco Bonaccorso*

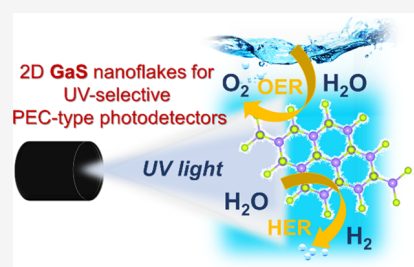

**ABSTRACT:** Two-dimensional (2D) transition-metal monochalcogenides have been recently predicted to be potential photo(electro)catalysts for water splitting and photoelectrochemical (PEC) reactions. Differently from the most established InSe, GaSe, GeSe, and many other monochalcogenides, bulk GaS has a large band gap of ∼2.5 eV, which increases up to more than 3.0 eV with decreasing its thickness due to quantum confinement effects. Therefore, 2D GaS fills the void between 2D small-band-gap semiconductors and insulators, resulting of interest for the realization of van der Waals type-I heterojunctions in photocatalysis, as well as the development of UV light-emitting diodes, quantum wells, and other optoelectronic devices. Based on theoretical calculations of the electronic structure of GaS as a function of layer number reported in the literature, we experimentally demonstrate, for the first time, the PEC properties of liquid-phase exfoliated GaS nanoflakes. Our results indicate that solution-processed 2D GaS-based PEC-type photodetectors outperform the corresponding solid-state photodetectors. In fact, the 2D morphology of the GaS flakes intrinsically minimizes the distance between the photogenerated charges and the surface area at which the redox reactions occur, limiting electron−hole recombination losses. The latter are instead deleterious for standard solid-state configurations. Consequently, PEC-type 2D GaS photodetectors display a relevant UV-selective photoresponse. In particular, they attain responsivities of 1.8 mA W$^{−1}$ in 1 M $H_2SO_4$ [at 0.8 V vs reversible hydrogen electrode (RHE)], 4.6 mA W$^{−1}$ in 1 M $Na_2SO_4$ (at 0.9 V vs RHE), and 6.8 mA W$^{−1}$ in 1 M KOH (at 1.1. V vs RHE) under 275 nm illumination wavelength with an intensity of 1.3 mW cm$^{−2}$. Beyond the photodetector application, 2D GaS-based PEC-type devices may find application in tandem solar PEC cells in combination with other visible-sensitive low-band-gap materials, including transition-metal monochalcogenides recently established for PEC solar energy conversion applications.

## INTRODUCTION

Gallium sulfide (GaS) is a binary IIIA−VIA group compound, which has gained increasing attention among the plethora of layered semiconductors due to its distinctive optoelectronic and anisotropic structural properties.[1−3] Depending on the stacking of layers, four GaS polytypes ($β$, $ε$, $γ$, and $δ$) are distinguished,[4] although the hexagonal (2H phase) $β$-polytype[5,6] is the most energetically favorable crystal arrangement.[4] A single layer of $β$-GaS is composed of S−Ga−Ga−S repeating units, with different layers kept together along the $c$-axis by weak van der Waals forces.[7,8] Differently from other investigated transition-metal monochalcogenides (e.g., GaSe,[9,10] InSe,[11,12] GeSe,[13] and SnSe[14]), the bulk form of GaS has a large optical band gap ($E_g$) (at 300 K: indirect $E_g$ ∼ 2.5 eV;[15−18] direct $E_g$ ∼ 3.0 eV[17−19]). The $E_g$ drastically increases above 3 eV with decreasing thickness down to the monolayer state due to quantum confinement effects.[20,21] Therefore, two-dimensional (2D) GaS fills the void between 2D small-$E_g$ semiconductors and insulators, which is of interest for the realization of ultraviolet (UV)-selective photodetectors,[22−24] color-tuneable blue/UV light-emitting diodes (LEDs),[20] and van der Waals type-I heterojunctions in photocatalysis.[21,25−27] Meanwhile, 2D GaS emerged as a potential material for applications such as electrochemical water splitting,[28] hydrogen storage,[29] energy storage (e.g., Li-ion batteries),[30,31] gas sensing,[32,33] DNA sequencing,[34] and nonlinear optics.[35,36] Contrary to several 2D materials, which are reactive to air (e.g., elemental analogue of graphene, such as silicene, germanene, and stanene[37,38] as well as transition-metal tellurides[39,40]) or undergo photoinduced oxidation (e.g., phosphorene[41−43] and metal monochalcogenides such as GaSe[44−47] and GeSe[48]), nearly ideally stoichiometric 2D GaS is oxidation-resistant under both laser/strong UV illumination[22] and mechanical stress,[49,50] showing a high

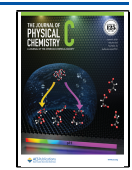

11857

https://doi.org/10.1021/acs.jpcc.1c03597
*J. Phys. Chem. C* 2021, 125, 11857−11866

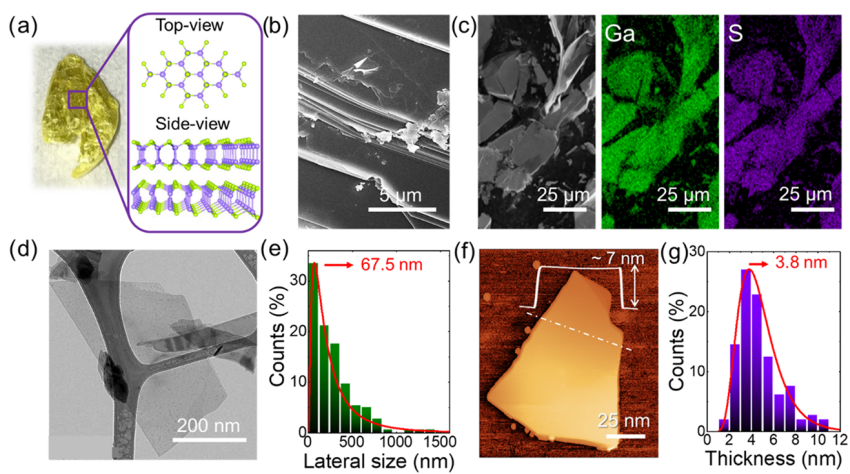

**Figure 1.** (a) Photograph of a β-GaS crystal synthesized through the direct reaction of Ga and S elements. The 2H structure ($P6_3/mmc$) of the crystal polytype is also shown. (b) SEM image of a fragment of the GaS crystal, showing its layered structure. (c) SEM image of fragments of GaS crystals and the corresponding EDS maps for Ga (Kα = 9.3 keV, green) and S (Kα = 2.3 keV, violet). (d) Bright-field transmission electron microscopy (BF-TEM) image of representative GaS flakes produced by the LPE of fragmentized GaS crystals. (e) BF-TEM statistical analysis of the lateral dimension of the GaS flakes. (f) Atomic force microscopy (AFM) image of a representative GaS flake. The height profile of a flake section is also shown. (g) AFM statistical analysis of the thickness of the GaS flakes.

activation energy (~3.1 eV) for the dissociation and chemisorption of $O_2$ molecules.[51]

Despite the appealing features of 2D GaS, its optoelectronic properties have been mainly investigated in solid-state photodetectors based on isolated flakes,[22,23,32] which were produced through mechanical cleavage[23,32] or chemical vapor deposition.[22] However, these techniques suffer from intrinsic scalability limits for their use in massive applications.[52,53] Liquid-phase exfoliation (LPE) methods can provide scalable production of 2D materials in the form of a liquid dispersion,[53–55] enabling their processing into thin films through low-cost and scalable deposition techniques,[12,56–58] including roll-to-roll printing.[59] Although the LPE represents a viable approach to exploit 2D GaS,[28,33] the printing of these materials produces percolating networks of flakes, which can inevitably lead to poor performances of optoelectronic devices compared to those measured for isolated flakes.[12,60] This effect is ascribed to the high contact resistance between flakes compared to the intrinsic resistance of the flakes themselves.[12,60–63] Therefore, it is pivotal to provide a paradigm shift in the design of printed optoelectronic devices to fully exploit the unique properties of solution-processed GaS flake films. In this context, ground-breaking experimental works demonstrated that group-IIIA monochalcogenides, including GaS, display electrochemical activities toward the hydrogen evolution reaction (HER), even though at high overpotentials (typically >0.4 V).[11,28] Meanwhile, their 2D forms have emerged as potential photo(electro)catalysts for both water splitting reactions, i.e., HER and oxygen evolution reaction (OER),[9,21,27] allowing new types of photoelectrochemical (PEC)-type photodetectors to be designed.[9] Importantly, the number of their layers controls the energy of the conduction band minimum and valence band maximum ($E_{CBM}$ and $E_{VBM}$, respectively). Consequently, group-IIIA monochalcogenides can be engineered to fulfill the fundamental requirements for the water splitting photo(electro)catalysts,[13,21,27] i.e., (1) $E_{CBM}$ > reduction energy level of $H^+/H_2$ ($E(H^+/H_2)$) and (2) $E_{VBM}$ < reduction energy level of $O_2/H_2O$ ($E(O_2/H_2O)$).[64–66] Despite the existence of few experimental studies on the PEC properties of the most established monochalcogenides, such as InSe,[67] GaSe,[9] and GeSe,[13] no PEC characterizations have been reported for GaS, which is, therefore, a subject matter of interest for the realization of UV-harvesting components in photocatalytic tandem structures (e.g., van der Waals type-I heterojunctions). In this context, the double peak feature around the Γ-point in the so-called "Mexican-hat-like" ring-shaped valence band dispersion of single-/few (≤5)-layer GaS flakes effectively enhances the photoabsorption cross section. In fact, the electrons available for the optical transitions are twice those available in the case of a single-peak valence band minimum.[23,27,68] Thanks to the 2D nature of GaS flakes, electrons and holes are directly photogenerated at the interface with the electrolyte, where redox reactions take place before the charges recombine.[69–72] This last feature avoids the need for high-mobility active materials, which are instead mandatory for high-responsivity solid-state photodetectors.[70,73,74]

By rationalizing the above observations, we report for the first time the use of 2D GaS flakes for the realization of quasi-visible-blind UV-selective PEC-type photodetectors operating in aqueous media.

## METHODS

The methods concerning materials, synthesis, and exfoliation of GaS crystals, materials characterization, fabrication of the photoelectrochemical (PEC)-type and solid-state photodetectors, and characterization of the photodetectors are reported in the Supporting Information (SI).

## RESULTS AND DISCUSSION

**Materials Characterization.** GaS single-/few-layer flakes were produced through ultrasonication-assisted LPE of bulk β-polytype in eco-friendly anhydrous 2-propanol (IPA) (see SI, Methods section).[75] The use of IPA as the exfoliating solvent circumvents the processability drawbacks related to the use of high-boiling-point and toxic solvents often used for the exfoliation of layered materials,[76–78] e.g., N-methyl-2-pyrrolidone (NMP) for graphene[79,80] and several metal chalcogenides.[81] Figure 1a shows a photograph of a representative GaS crystal, together with the top- and side-views of its hexagonal double-layered structure with the Ga–Ga and Ga–S distances of 2.48 and 2.37 Å, respectively,[82,83] and an interlayer distance approximating the monolayer thickness of ∼0.75 nm.[83] Figure 1b shows the scanning electron microscopy (SEM) image of a fragment of the GaS crystal, evidencing its layered structure. The SEM-coupled energy-dispersive X-ray spectroscopy (EDS) analysis (Figure 1c and Table S1) indicates a nearly ideal stoichiometry of the GaS crystal (Ga-to-S atomic ratio ∼ 1.05).

Figure 1d shows the bright-field transmission electron microscopy (BF-TEM) image of a representative LPE-produced GaS flake, which displays a nearly rectangular shape with regular edges. The lateral sizes of the flakes follow a log-normal distribution peaked at 67.5 nm (Figure 1e). Figure 1f shows the atomic force microscopy (AFM) image of a representative exfoliated flake together with its height profile, corresponding to a thickness of ∼7 nm. Figure S1 shows another AFM image displaying few-layer GaS flakes with thickness ranging between 1.9 and 4.3 nm. The statistical analysis of the thickness data reports values ranging from 1.5 to 12 nm, whose distribution is fitted by a log-normal curve peaked at 3.8 nm (Figure 1g). The experimental AFM thickness of monolayer GaS being between 0.85 and 1 nm,[20,22,84] close to the theoretical value of 0.75 nm,[83,84] our GaS flakes mainly consist of few (≤5)-layer flakes.

Figure 2a shows the optical extinction spectrum of the LPE-produced GaS flake dispersion. The plot monotonically increases with decreasing wavelength until a narrow peak around ∼280 nm is reached. The tail of the spectrum in the visible and near-IR spectral region has been previously ascribed to the scattering contribution of dispersed flakes with nanoscale dimensions.[28] The concentration of the as-produced GaS flake dispersion was first measured by weighing the solid material content in a known volume of dispersion, giving a value of ∼0.2 g L$^{-1}$. The extinction coefficient of the GaS flakes was estimated using the Lambert–Beer law, Ext($\lambda$) = $\varepsilon(\lambda)cl$, in which $\lambda$ is a given optical wavelength, Ext($\lambda$) is the optical extinction at $\lambda$, $\varepsilon(\lambda)$ is the extinction coefficient at $\lambda$, $c$ is the material concentration, and $l$ is the optical path length.[85] By measuring the optical extinction spectra of the as-produced GaS flake dispersions with controlled concentrations, $\varepsilon$(285 nm) is found to be ∼199 L g$^{-1}$ m$^{-1}$. To guarantee the reproducibility of the material deposition processes, the concentration among different batches of GaS flake dispersion has been finely controlled by monitoring their optical extinction spectrum, i.e., $c$ = Ext($\lambda$)/($\varepsilon(\lambda)l$). The $E_g$ of the GaS flakes was determined by diffuse reflectance spectroscopy (DRS) of a film of GaS flakes, deposited through spray coating, using the Kubelka–Munk theory of diffuse reflectance ($R$).[86,87] In particular, the $E_g$ can be estimated by fitting the linear part of $(F(R)h\nu)^n$ vs $h\nu$ (Tauc plot) with $(F(R)h\nu)^n = Y(h\nu - E_g)$ (Tauc relation), in which $F(R)$ is the Kubelka–Munk function, defined as $F(R) = (1 - R)^2/2R$, $h$ is Planck's constant, $\nu$ is the photon's frequency, and $Y$ is a proportionality constant.[86,87] The value of $n$ specifies the type of the electronic transitions, distinguishing between direct ($n$ = 2) and indirect interband transitions ($n$ = 0.5).[88–90] Figure 2b shows the Tauc plots of the GaS flake film for both $n$ = 2 and 0.5. The estimated direct $E_g$ is ∼2.9 eV, while the indirect $E_g$ is ∼2.6 eV. These $E_g$ values agree with those reported in the literature for few-layer GaS flakes.[20] It is noteworthy that the sprayed GaS-based films consist of flakes with polydisperse morphological features. Therefore, the energy onset referring to $E_g$ of the thickest nanoflakes may experimentally hide those of the thinnest ones, since the latter show higher $E_g$ (theoretical values > 3 eV[21,27]).[20] Figure 2c reports the X-ray diffraction (XRD) pattern of the GaS flakes in comparison with the one measured for GaS crystal powder. The XRD peaks of the exfoliated sample resemble those of the bulk crystals, which are indexed to the 2H structure of $\beta$-GaS (ICSD-173940) with lattice parameters $a = b$ = 3.627 Å and $c$ = 17.425 Å.[91] The absence of characteristic peaks attributed to crystalline impurities, such as Ga$_2$O$_3$, indicates that both the synthesis and the subsequent LPE of GaS crystals generate products with marginal defects. The structural properties of the bulk and exfoliated GaS crystals were further evaluated by Raman spectroscopy measurements. The group theory predicts six nondegenerate Raman active optical modes for the $P6_3/mmc$ ($D_{6h}$) space group of bulk $\beta$-GaS, i.e., 2A$_{1g}$ + 2E$_{1g}$ + 2E$_{2g}$.[92–94] The most intense and investigated ones are A$_{1g}^1$, A$_{1g}^2$, and E$_{2g}^1$ (the latter often includes the contribution of the nearby E$_{1g}^2$).[92,93] These modes are also observed in the exfoliated GaS,[94,95] and their intensities decrease with the reduction of the number of layers.[22,94] In particular, for the noncentrosymmetric monolayer GaS (space group: $D_{3h}$), E$_{2g}^1$ is typically indistinguishable from the Raman signal of the Si substrate.[22,94] Recent studies demonstrated that the peak position of A$_{1g}^1$ is a trustworthy indicator for simple and fast determination of the thickness of the GaS flakes.[22,94] In fact, the A$_{1g}^1$ peak is softened (red-shifted) following a decrease of the number of layers due to the reduced impact of the interlayer interaction on phonon restoring forces.[22,94] This phenomenon has also been detected in GaSe[9] and other transition-metal dichalcogenides (e.g., MoS$_2$).[76,96] Figure 2d shows the Raman spectra of both bulk and exfoliated GaS crystals measured with an excitation wavelength ($\lambda_{exc}$) of 514 nm. In agreement with the above consideration, the A$_{1g}^1$ peak position for the GaS flakes is slightly red-shifted compared to the bulk case, indicating the successful exfoliation of the crystals through the LPE method. The quantitative statistical analysis of the $A_{1g}^1$ peak position (calculated on 40 different spectra) is reported in Figure S2. In addition, Raman spectra of the GaS flakes do not exhibit characteristic peaks attributed to other crystalline species beyond GaS (e.g., $Ga_2O_3$, showing a pronounced Raman mode peak at ∼200 $cm^{−1}$).[97] Therefore, these results further support that the LPE process of GaS crystals carried out in IPA does not cause relevant oxidation effects, in agreement with the XRD measurements and the X-ray photoelectron spectroscopy (XPS) analysis (Figures S3 and S4). High-resolution TEM (HRTEM) analysis was performed to further examine the crystal structure of the LPE-produced GaS flakes. Figure 3a shows a BF-TEM image of a portion of a representative GaS flake near its edge. The corresponding selected area electron diffraction (SAED) pattern (inset of Figure 3a) matches that of the 2H structure of β-GaS (ICSD-173940), in agreement with the XRD analysis.

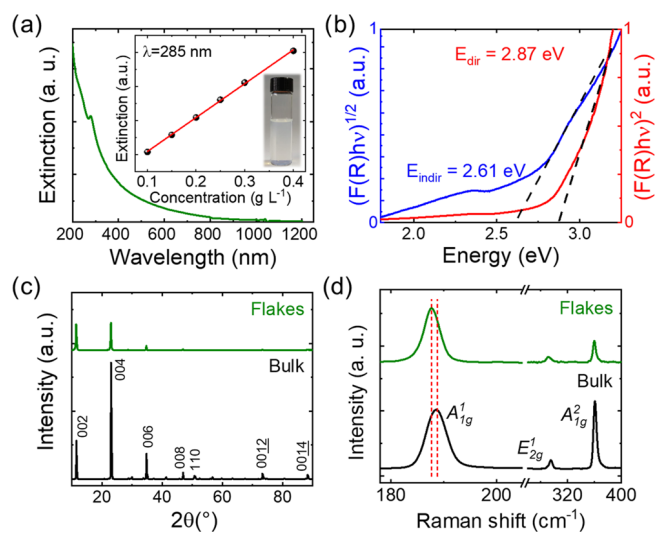

**Figure 2.** (a) Optical extinction spectrum of the LPE-produced GaS flake dispersion. The inset shows the Ext(285 nm) vs $c$, together with a photograph of a GaS flake dispersion. (b) $(F(R)h\nu)^n$ vs $h\nu$ (Tauc plots) measured for the GaS flakes for direct ($n$ = 2, red trace) and indirect ($n$ = 0.5, blue trace) interband transitions. (c) XRD patterns and (d) Raman spectra ($\lambda_{exc}$ = 514 nm) of bulk (black curve) and exfoliated (green curve) GaS crystals. Panels (c) and (d) report the diffraction peaks and Raman modes attributed to the 2H structure of $\beta$-GaS, respectively.

Figure 3b shows the HRTEM image of a portion of the GaS flake, confirming the lattice spacings of the β-GaS. Scanning transmission electron microscopy coupled with EDS (STEM-EDS) analyses were carried out to evaluate the composition of the GaS flakes. Figure 3c shows the STEM image of partially suspended GaS flakes. Figure 3d shows the corresponding STEM-EDS maps of Ga, S, O, and C. The quantitative elemental analysis results in a S-to-Ga atomic ratio of ∼1.1 and a low atomic content of O (O-to-Ga atomic ratio ∼ 0.08), which excludes the relevant presence of oxide domains near the edges of the flakes.

**Photoelectrochemical Characterization.** The LPE-produced GaS nanoflake dispersions were sprayed onto graphite paper, which acts as a non-photoactive current collector, to produce photoelectrode films. The PEC properties of the as-produced devices were evaluated in a three-electrode system (Figure 4a) in different aqueous electrolytes: acidic 1 M $H_2SO_4$ (pH = 0.5), near-neutral 1 M $Na_2SO_4$ (pH = 6), and alkaline 1 M KOH (pH = 14), investigating the response under UV/visible excitation wavelengths. To the best of our knowledge, the PEC properties of GaS flakes are currently unknown, although theoretical studies predicted their potential as water splitting photocatalysts.[21,27] It is important to underline that the production of PEC-type devices is generally advantageous in terms of cost and ease of implementation compared to photodetectors based on isolated flakes.[98−100] In principle, solution-processed 2D material films can also be deposited on interdigitated electrode to produce solid-state photodetectors. However, differently from solid-state photodetectors based on other solution-processed films of transition metal monochalcogenides (such as InSe),[12,60] our attempts on photodetectors based on sprayed GaS flake films have shown poor device performance, i.e., responsivity lower than 0.4 mA $W^{−1}$ under 275 nm illumination with an intensity of 1.3 mW $cm^{−2}$ (see Optical Characterization in the Supporting Information, Figure S5). These results are ascribed to the low mobility of the GaS flakes (∼0.1 $cm^2$ $V^{−1}$ $s^{−1}$),[61] which results in a highly resistive percolating network hindering the immediate exploitation of the peculiar optoelectronic proper-ties of the solution-processed 2D GaS. Figure 4b reports a photograph of the GaS photoelectrode. Figures 4c and S6 show top-view SEM images of a GaS photoelectrode, which displays a porous film of flakes preferentially oriented with planes mostly parallel to the current collector.

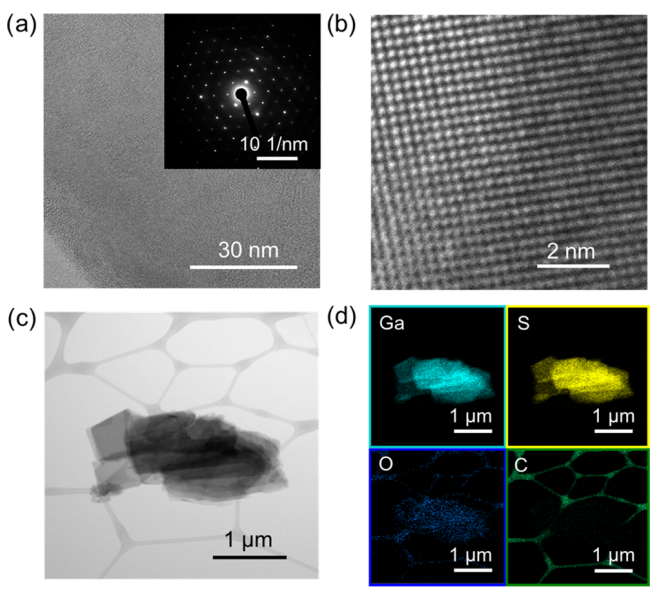

Figure 3. (a) BF-TEM image of a portion of a representative LPE-produced GaS flake near its edge. The inset shows the corresponding SAED pattern, which matches that of the 2H structure of β-GaS. (b) HRTEM image of a portion of the GaS flakes. (c) Scanning transmission electron microscopy (STEM) image of a partially suspended GaS flake. (d) Corresponding quantitative STEM-EDS maps of Ga (Kα1), S (Kα1), O (Kα1), and C (Kα1_2).

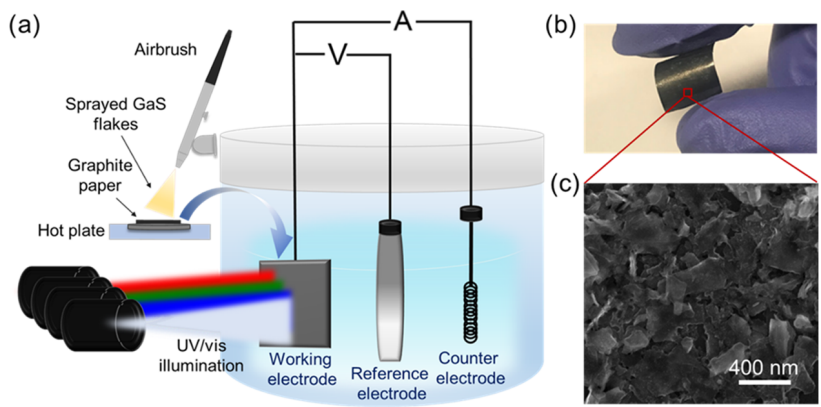

Figure 4. (a) Sketch of the experimental setup used for characterization of the PEC-type GaS photoelectrodes, which were produced by spray coating the GaS nanoflakes on a substrate of graphite paper, acting as the current collector. (b) Photograph of a flexible GaS photoelectrode. (c) Top-view SEM image of a GaS photoelectrode.

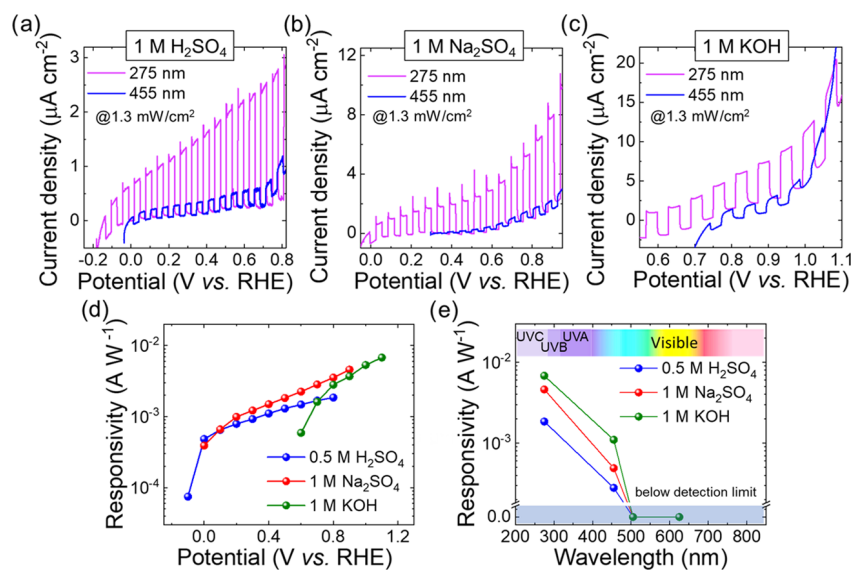

**Figure 5.** LSV scans of GaS PEC-type photodetectors in (a) 1 M $H_2SO_4$, (b) 1 M $Na_2SO_4$, and (c) 1 M KOH under UV (275 nm) and blue (455 nm) illumination with an intensity of 1.3 mW cm$^{-2}$. (d) Potential dependence of the responsivity of the GaS PEC-type photodetectors under 275 nm illumination with an intensity of 1.3 mW cm$^{-2}$ in the three investigated aqueous media. For the device operating in 1 M KOH, the potentials at which the device displayed significant negative dark current were excluded from the x-axis data range. (e) Wavelength dependence of the device responsivity under the same illumination intensity of 1.3 mW cm$^{-2}$ in the three investigated aqueous media.

The GaS photoelectrodes were evaluated as PEC-type photodetectors for four different illumination wavelengths in the UV/visible spectral range, namely, 275, 455, 505, and 625 nm. It is noteworthy that only the illumination wavelength of 275 nm (hard UV−UVC region) corresponds to energy above the direct $E_g$ estimated for the GaS flakes (∼2.9 eV), while both 275 and 455 nm correspond to energy higher than the material indirect $E_g$ (∼2.6 eV). Figure 5a−c shows the anodic linear sweep voltammetry (LSV) measurements for the GaS photodetectors under chopped illumination (frequency = 0.33 Hz) at excitation wavelengths of 275 and 455 nm with an intensity of 1.3 mW cm$^{-2}$ in the three investigated aqueous media, i.e., 1 M $H_2SO_4$, 1 M $Na_2SO_4$, and 1 M KOH. It is noteworthy that previous first-principles calculations predicted a photocatalytic activity of GaS flakes toward the OER.[27] In detail, despite the dependence of the $E_g$ on the number of layers, it was proven that both single- and few- (≤5)-layer GaS flakes, as well as multi (>5)-layer GaS flakes, fulfill the fundamental requirements to carry out the OER, i.e., $E_{VBM}$ < $E(O_2/H_2O)$ for all of the pH conditions investigated in our work.[27] Meanwhile, our preliminary cathodic LSV measurements did not show any photocathodic response of the GaS photoelectrodes, which were therefore analyzed only as photoanodes. In acidic media, the absence of the photocathodic response is attributed to the poor electrocatalytic activity of few-layer GaS flakes toward the HER (experimental reaction overpotential higher than 0.4 V).[28] In addition, in mild acidic and neutral or alkaline conditions, the energy offset between the $E_{CBM}$ and $E(H^+/H_2)$ is significantly smaller than the one between $E_{VBM}$ and $E(O_2/H_2O)$.[27] Therefore, the electronic structure of 2D GaS promotes the OER in a PEC cell architecture without any charge extraction layers and/or cocatalysts, as in our case. It is noteworthy that an anodic PEC behavior similar to our GaS flakes has also been observed for 2D InSe,[67] for which an unsatisfactory electrocatalytic activity toward HER has been observed in both acidic and alkaline media.[101] To avoid the electrochemically induced degradation of the photoelectrode, the applied potentials were limited to regions with dark current density substantially inferior to the detected photocurrents. Moreover, zero (or below-detection sensitivity) photoresponses were detected for the illumination wavelengths of 505 and 625 nm. In all of the investigated media, the photoresponses of the photoanodes under UV light (275 nm) are significantly higher than those measured for blue light (455 nm), indicating a UV-selective photoresponse. Moreover, in 1 M KOH, the photoanodes displayed negative dark currents above the open-circuit potential under illumination, i.e., at a potential between 0.3 and 0.8 V vs reversible hydrogen electrode (RHE). Since the chemical reactivity of the substrate was excluded for such conditions, the negative dark current at a potential inferior to ∼0.8 V vs RHE can be likely ascribed to corrosion effects involving GaS flakes.

Figure 5d reports the responsivity of the GaS photoanodes as a function of the applied potential in the three investigated media. The highest recorded responsivities for the 275 nm illumination are 1.8 mA W$^{-1}$ for 1 M $H_2SO_4$ at 0.8 V vs RHE, 4.6 mA W$^{-1}$ for 1 M $Na_2SO_4$ at 0.9 V vs RHE, and 6.8 mA W$^{-1}$ for 1 M KOH at 1.1. V vs RHE. Importantly, the highest photoresponse of GaS flakes has been observed in alkaline conditions, since the energy offset between $E_{VBM}$ and $E(O_2/H_2O)$ is maximized compared to both the neutral and acidic media.[27] Figure 5e shows the wavelength-dependent responsivity of the GaS photodetectors, indicating UV-selective light detection for UV-sensitive applications. These results agree with previous studies on solid-state photodetectors based on stoichiometric GaS flakes.[22] Overall, these results suggest that the direct band gap transition is the main pathway driving the PEC activity of the devices. The low responsivity to blue light (455 nm) can be likely ascribed to either indirect band gap absorption or sub-band-gap states, which can also contribute to the photoresponse to blue light in solid-state photodetectors based on isolated GaS flakes.[22] It is noteworthy that indirect band gap transitions also involve phonons, resulting in ineffective light absorption, especially in ultrathin films.[102] Table S2 shows the comparison between the responsivity of our PEC-type GaS photodetectors and those of other solution-processed UV photodetectors based on 2D materials. Clearly, the combination of GaS flakes and the use of a PEC-type architecture is effective at overcoming the performance of several solid-state UV photodetectors based on solution-processed 2D materials reported in previous literature.[103−107] In addition, the performances of our PEC-type GaS photodetectors can be one order of magnitude higher than those exhibited by their solid-state analogues based on GaS flake films deposited onto interdigitated electrodes (Figure S4). Lastly, the validation of the GaS-based PEC-type photodetectors can be useful for the design of PEC analytic systems that operate with low voltage sources,[67,108,109] without recurring complex device manufacturing. For example, the detection of an analyte by means of PEC reactions can offer several benefits compared to electrochemical sensors. For example, PEC-type sensors can reduce the background signal down to the limit of lock-in detection noise when they operate in differential mode.[110] Meanwhile, contrary to electrochemical sensors, they can eliminate the need for numerous recalibrations.[111,112] Therefore, our validation of PEC-type GaS-based photodetectors with UV sensitivity may pave a new way for the realization of novel concepts of sensing devices.

The stability of GaS photodetectors was evaluated through subsequent LSV scans. As shown in Figure S7a, the GaS photoelectrodes show Raman spectra similar to the one measured for the as-produced GaS flakes, which means that the GaS flakes retain their starting structural properties during the PEC tests. The devices exhibit the most stable PEC performance in 1 M KOH, showing a progressive responsivity stabilization over the first 10 LSV scans (Figure S7b). The initial degradation may be ascribed to the mechanical delamination induced by progressive gas evolution (i.e., oxygen evolution due to OER), as previously reported for similar architectures based on transition-metal monochalcogenides (e.g., GaSe[9] and GeSe[13]). Prospectively, the use of polymeric (e.g., Nafion)[13,113,114] or conductive (e.g., carbon nanotubes) binders,[115−117] as well as the engineering of devices with effective charge-extracting layers and cocatalysts,[118−120] could help to further stabilize the GaS photoelectrodes.

## CONCLUSIONS

In summary, we presented the first PEC characterization of 2D GaS flakes produced through LPE methods in IPA. Our results provide novel insights into the fundamental PEC properties of 2D GaS flakes, which can be used to design and realize innovative PEC-type UV-selective photodetectors for medical diagnostics, air purification, chemical analysis (ozone sensing), and advanced optical communication systems.[121−125] In particular, LPE-produced GaS flakes can be easily deposited through printing techniques to produce solution-processed photocatalytic films on graphite paper, the latter acting as a current collector of the resulting photoelectrodes. Our PEC characterizations indicate that solution-processed 2D GaS-based PEC-type photodetectors outperform the corresponding solid-state photodetectors. In fact, the 2D morphology of the GaS flakes intrinsically minimizes the distance between the photogenerated charges and the surface area at which the redox reactions occur, limiting electron−hole recombination losses compared to the case of standard solid-state configurations made of the same photoactive films. Therefore, our PEC-type GaS photodetectors display a relevant UV-selective PEC photoresponse, attaining responsivities of 1.8 mA W$^{-1}$ in 1 M H$_2$SO$_4$ (at 0.8 V vs RHE), 4.6 mA W$^{-1}$ in 1 M Na$_2$SO$_4$ (at 0.9 V vs RHE), and 6.8 mA W$^{-1}$ in 1 M KOH (at 1.1. V vs RHE) under 275 nm illumination wavelength with an intensity of 1.3 mW cm$^{-2}$. Beyond the photodetector application, GaS-based PEC-type devices can find application in tandem solar PEC cells in combination with other visible-sensitive low-band-gap materials, including transition-metal monochalcogenides recently established for PEC solar energy conversion applications.[126−128]

## ■ ASSOCIATED CONTENT

### ⓈSupporting Information

The Supporting Information is available free of charge at https://pubs.acs.org/doi/10.1021/acs.jpcc.1c03597.

Methods, AFM analysis, SEM−EDS analysis, Raman spectroscopy analysis, XPS characterization, characterization of the solid-state photodetectors, and performance comparison with the literature (PDF).

## ■ AUTHOR INFORMATION

### Corresponding Authors

**Sebastiano Bellani** − BeDimensional Spa., 16163 Genova, Italy; Graphene Labs, Istituto Italiano di Tecnologia, 16163 Genova, Italy; Email: s.bellani@bedimensional.it

**Francesco Bonaccorso** − BeDimensional Spa., 16163 Genova, Italy; Graphene Labs, Istituto Italiano di Tecnologia, 16163 Genova, Italy; orcid.org/0000-0001-7238-9420; Email: f.bonaccorso@bedimensional.it

### Authors

**Marilena I. Zappia** − BeDimensional Spa., 16163 Genova, Italy; Department of Physics, University of Calabria, 87036 Rende, CS, Italy

**Gabriele Bianca** − Graphene Labs, Istituto Italiano di Tecnologia, 16163 Genova, Italy; Dipartimento di Chimica e Chimica Industriale, Università degli Studi di Genova, 16146 Genoa, Italy

**Nicola Curreli** − Functional Nanosystems, Istituto Italiano di Tecnologia (IIT), 16163 Genova, Italy

**Zdeněk Sofer** − Department of Inorganic Chemistry, University of Chemistry and Technology Prague, 166 28 Prague 6, Czech Republic; orcid.org/0000-0002-1391-4448

**Michele Serri** − Graphene Labs, Istituto Italiano di Tecnologia, 16163 Genova, Italy; orcid.org/0000-0002-6018-5284

**Leyla Najafi** − BeDimensional Spa., 16163 Genova, Italy; Graphene Labs, Istituto Italiano di Tecnologia, 16163 Genova, Italy

**Marco Piccinni** − Graphene Labs, Istituto Italiano di Tecnologia, 16163 Genova, Italy; Dipartimento di Chimica e Chimica Industriale, Università degli Studi di Genova, 16146 Genoa, Italy

**Reinier Oropesa-Nuñez** − BeDimensional Spa., 16163 Genova, Italy; Department of Material Science and

Engineering, Uppsala University, 75121 Uppsala, Sweden; orcid.org/0000-0002-9551-6565

**Petr Marvan** − *Department of Inorganic Chemistry, University of Chemistry and Technology Prague, 166 28 Prague 6, Czech Republic*

**Vittorio Pellegrini** − *BeDimensional Spa., 16163 Genova, Italy; Graphene Labs, Istituto Italiano di Tecnologia, 16163 Genova, Italy*

**Ilka Kriegel** − *Functional Nanosystems, Istituto Italiano di Tecnologia (IIT), 16163 Genova, Italy;* orcid.org/0000-0002-0221-3769

**Mirko Prato** − *Materials Characterization Facility, Istituto Italiano di Tecnologia, Genova 16163, Italy;* orcid.org/0000-0002-2188-8059

**Anna Cupolillo** − *Department of Physics, University of Calabria, 87036 Rende, CS, Italy*

Complete contact information is available at:
https://pubs.acs.org/10.1021/acs.jpcc.1c03597

### Author Contributions

M.I.Z. and G.B. contributed equally to this work. The manuscript was written through contributions of all authors. All authors have given approval to the final version of the manuscript.

### Notes

The authors declare no competing financial interest.

## ■ ACKNOWLEDGMENTS

This project has received funding from the European Union's Horizon 2020 European Research Council under grant agreement no. [850875] (Light-DYNAMO) and from the European Union's Horizon 2020 Research and Innovation programme under grant agreement no. [101017821] (LIGHT-CAP) and grant agreement no. 881603-GrapheneCore3, the MSCA-ITN ULTIMATE project under grant agreement no. 813036, and the Bilateral project GINSENG between NSFC (China) and MAECI (Italy) (2018−2020) by the Natural Science Foundation of Shandong Province (ZR2019QEM009). This project was supported by the Czech Science Foundation (GACR No. 20-16124J). P.M. was supported by specific university research grant no. A2_FCHT_2020_055. M.I.Z. received funding from PON Research and Innovation 2014−2020 (CUP H25D18000230006) by the Italian Ministry of University and Research. We thank the Materials Characterization Facility—Istituto Italiano di Tecnologia—for the support in XRD data acquisition/analysis and Electron Microscopy facility—Istituto Italiano di Tecnologia—for the support in TEM data acquisition.

## ■ ABBREVIATIONS

2D, two-dimensional; AFM, atomic force microscopy; EDS, energy-dispersive X-ray spectroscopy; $E_g$, optical band gap; Ga, gallium sulfide; IPA, 2-propanol; LED, light-emitting diode; LPE, liquid-phase exfoliation; HER, hydrogen evolution reaction; NMP, N-methyl-2-pyrrolidone; $R$, diffusive reflectance; RHE, reversible hydrogen electrode; OER, oxygen evolution reaction; PEC, photoelectrochemical; SEM, scanning electron microscopy; TEM, transmission electron microscopy; UV, ultraviolet; XPS, X-ray photoelectron spectroscopy; XRD, X-ray diffraction